\documentclass[sigplan,screen]{acmart}
\usepackage{graphicx} 

\setcopyright{rightsretained}
\acmDOI{10.1145/3689492.3689816}
\acmYear{2024}
\copyrightyear{2024}
\acmISBN{979-8-4007-1215-9/24/10}
\acmConference[Onward! '24]{Proceedings of the 2024 ACM SIGPLAN International Symposium on New Ideas, New Paradigms, and Reflections on Programming and Software}{October 23--25, 2024}{Pasadena, CA, USA}
\acmBooktitle{Proceedings of the 2024 ACM SIGPLAN International Symposium on New Ideas, New Paradigms, and Reflections on Programming and Software (Onward! '24), October 23--25, 2024, Pasadena, CA, USA}
\acmSubmissionID{onward24essays-p18-p}
\received{2024-04-25}
\received[accepted]{2024-08-08}


\title{tl;dr: Chill, y'all: AI Will Not Devour SE}

\author{Eunsuk Kang}
\orcid{0000-0001-7891-6885}
\affiliation{%
  \institution{Carnegie Mellon University}
  \city{Pittsburgh}
  \country{USA}
}
\email{eunsukk@andrew.cmu.edu}

\author{Mary Shaw}
\orcid{0000-0003-1337-4557}
\affiliation{%
  \institution{Carnegie Mellon University}
  \city{Pittsburgh}
  \country{USA}
}
\email{mary.shaw@cs.cmu.edu}


\date{April 2024}

\begin{document}

\begin{abstract}
 Social media provide a steady diet of dire warnings that artificial intelligence (AI) will make software engineering (SE) irrelevant or obsolete.  To the contrary, the engineering discipline of software is rich and robust; it encompasses the full scope of software design, development, deployment, and practical use; and it has regularly assimilated radical new offerings from AI. Current AI innovations such as machine learning, large language models (LLMs) and generative AI will offer new opportunities to extend the models and methods of SE. They may automate some routine development processes, and they will bring new kinds of components and architectures. If we're fortunate they may force SE to rethink what we mean by correctness and reliability. They will not,  however, render SE irrelevant.
\end{abstract}

\maketitle

\section{What’s with the cries of alarm about AI’s threats to SE?}

The blog-o-sphere and the X-formerly-twitter-sphere provide a regular stream of angst about the imminent demise of SE at the hands of AI.  It seems that either AI systems are so different from “regular” software systems that SE knowledge has become obsolete or irrelevant, or else AI will soon take over programming, and by extension software development.
\begin{itemize}
\item ``The End of Programming: The end of classical computer science is coming, and most of us are dinosaurs waiting for the meteor to hit''~\cite{Welsh23}.
\item ``NVIDIA CEO says the future of coding as a career might already be dead in the water with the imminent prevalence of AI. Generative AI could claim more jobs in the tech landscape, rendering coding professionals redundant''~\cite{Huang24}.
\item ``ChatGPT Will Replace Programmers Within 10 Years: Predicting The End of Manmade Software''~\cite{Hughes23}
\item ``Software engineers are panicking about being replaced by AI''~\cite{Kay23}
\end{itemize}

We’re here to say that those fears are overrated.  Software engineering is a rich, robust discipline that covers software systems from initial concept through their lifetime. SE has a long history of embracing and assimilating radical new ideas from AI. SE will, likewise, evolve to handle generative AI. SE does need to re-think its concept of correctness; generative AI may finally force SE to do this. How, then, should SE engage with generative AI?

\section{Software engineering is a rich, robust discipline that covers software systems from idea through their lifetime}

Engineering creates cost-effective solutions to practical problems by applying well-established knowledge to building things, in the service of mankind. It entails making decisions under constraints of limited time, knowledge, and resources.
\begin{quote}
Software engineering, then, is the branch of computer science that creates practical, cost-effective solutions to computing and information processing problems, by applying the best-systematized knowledge available, developing software systems in the service of mankind The distinctive character of  software raises special issues about its engineering, including:
\begin{itemize}
\item Software is design-intensive; manufacturing costs are a minor component of software product costs.
\item Software is symbolic, abstract, and more constrained by intellectual complexity than by fundamental physical laws~\cite{Manifesto}. 
\end{itemize}
\end{quote}

The discipline of software engineering encompasses the full scope of software systems from concept to retirement---a full spectrum of issues from understanding what problem the software should solve through overall design, tradeoff resolution, performance, reliability, sustainability, usability, fitness for purpose, programming of components, composition of components, validation, adherence to policy and standards, and evolution. 

Engineering design activities fall on a spectrum from routine to innovative design.  Routine design involves familiar problems, stereotypical definitions, and reusing lots of prior work. Innovative design, on the other hand, involves finding novel solutions to unfamiliar problems ~\cite{Prospects}. A real-world project usually involves a mix of these, and the innovative parts are the most resistant to automation. 

Software engineering suffers from some unfortunate misconceptions that arise from our origin myth, which says, “Software is created by professional programmers by writing code in sound programming languages to satisfy a given formal specification and verify that the program is correct; software is created by (just) composing program modules”~\cite{Myths}.

This origin myth has by now become quite a narrow view of software, but the mindset still persists. This leads to misconceptions such as that software engineering is simply programming\footnote{These two roles are often conflated, especially since programmers often have job titles such as “software engineer”.  In this essay, “programmers” primarily write code whose goals have been laid out by someone else; “engineers” are responsible for design, structure and quality attributes of a system (and “software engineers” do so for software systems). In the context of using generative AI to create software, “developers” use the AI tools to create software; when they are doing this directly at the level of code they resemble programmers.  “Vernacular developers” are people, often professionals in their own disciplines, who develop software for their own purposes but are not highly-trained computing professionals~\cite{Myths}; their needs and development techniques are often very different from programmers and software engineers. We prefer “vernacular developer” to “end user programmer” because it covers the full responsibility for the software and “vernacular” seems less dismissive than end user (“Only two industries call their customers `users'" ~\cite{pausch}).}, that all software should have specifications, and that most people creating software are trained professionals~\cite{Myths}. Even though SE  has  since moved beyond that origin myth, the mindset lingers in the form of emphasis on having at least informal specifications, on continued focus on correctness, and on poor support for vernacular developers.

\subsection{Progress in programming can be tracked by the increasing scale of abstractions used by programmers}

Let’s focus for a moment on the programming task, or rather the task of creating the code of a software system. Sixty and seventy years ago this was done largely at the level of the machine, in assembly language. We can mark progress since then by the increasing scale of the primitive elements programmers use. Languages such as Fortran and Cobol raised the level of abstraction in programs from the computer hardware to arithmetic expressions and elementary data structures. Then objects, filters, and other module-level constructs became the norm, along with notations for describing the composition of components~\cite{Abstraction}. 

The trend continues, and a good indicator of progress in programming is the increasing magnitude of the abstractions we treat as primitive---that is, the constructs that we generally use and trust as primitive elements, without looking inside. Software development now relies as heavily on composition of existing components as on writing lines of code. Indeed, there is currently considerable interest in “low-code” software development that enables inexperienced programmers to build apps from components through a visual interface.

\subsection{Software engineering interacts with corporate strategies and public policy}

SE encompasses not only the full spectrum of design and engineering decisions about software systems, but it is also embedded in the organizational context of software development and the way the products are embedded in society. It follows that both software engineers and programmers must understand that the software they develop is shaped by organization and policy considerations as well as the problem immediately at hand. 

This is evident in Curtis’ study of communication patterns in software development organizations. Curtis believed that “the design process had to be understood at many levels of analysis---cognitive, social, organizational, etc.,---in order to accurately characterize the problems experienced in developing large systems software. ... [our studies led to] ... a model of the different layers of human activity that provide the context for software development”~\cite{Curtisrainbow}.

\begin{figure}[!t]
\includegraphics[width=0.48\textwidth]{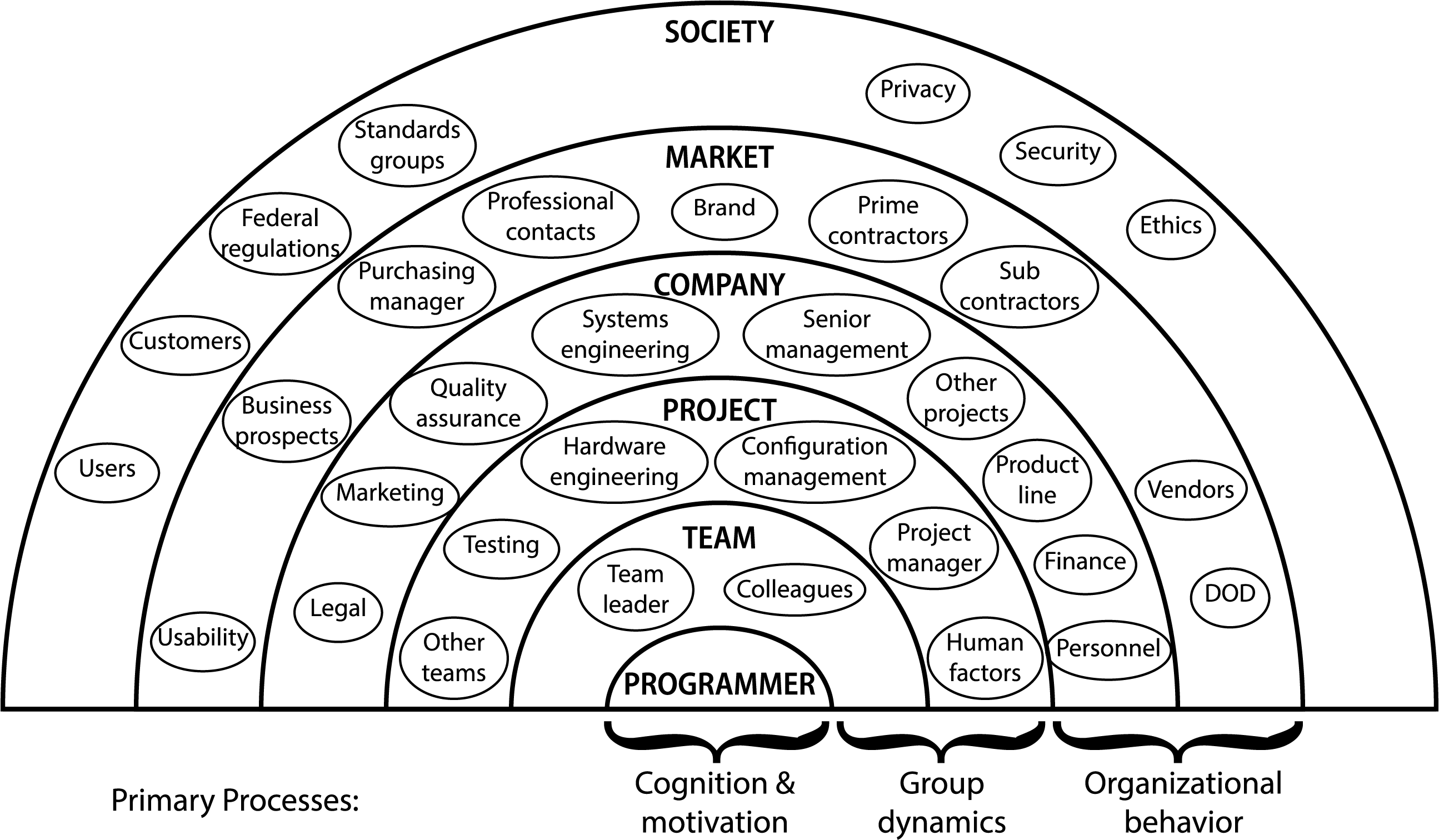}
\centering
\caption{Organizational context of software development (based on~\cite{Curtisrainbow} Figures 1 and 5)}
\label{fig:curtis-rainbow}
\end{figure}

Each higher level introduces new business and policy concerns that constrain the software design or the development process. Figure~\ref{fig:curtis-rainbow} shows some of the processes and concerns in each level. Each of these constrains the processes of inner levels. So, for example, programmers may be expected to observe company policies on documentation and style, and code details may be constrained by each jurisdiction’s privacy and security policies.

All computing professionals should be familiar with this organizational context and how it constrains their decisions. Indeed, ACM’s new undergraduate curriculum~\cite{ACM24} reaches into the outer rings of Figure~\ref{fig:curtis-rainbow}. It says that all computer science graduates should understand not only programming, teamwork, and team processes (including responsibilities and risks), but also responsible, collaborative professional behavior. It also calls for students specializing in SE to know how to interface with stakeholders including management, customers, and users; to elicit functional as well as quality attribute requirements; to identify risks and ethical considerations; and to use validation and reliability strategies appropriate to the application.

\subsection{Software engineering uses domain knowledge to bridge models of the world to models of the machine}

In his seminal paper~\cite{Jackson95}, Jackson introduces one of the most fundamental concepts in the engineering of software systems---the distinction between the world and the machine. Here, the machine corresponds to the software that will solve a given problem, and the world (or the domain) represents the part of the real world in which the machine is to be deployed. The machine interacts with the world through shared phenomena and has limited control over parts of the world that lie beyond this interface. Thus, to establish a requirement, which is expressed in terms of domain phenomena, it is not sufficient for the machine to satisfy its specification; the system must also rely on a set of assumptions about the domain. These assumptions are elicited and validated through consultation with domain experts and stakeholders in an early stage of development. If the requirement cannot be determined in advance, software developers may use software as a medium to investigate the assumptions of the domain via exploratory programming. 

For example, control software inside a self-driving vehicle observes parts of the physical world through a set of sensors (e.g., camera) and generates commands to manipulate an actuator (e.g., steering wheel). To demonstrate an acceptably low risk of collision, the system would need to assume that the sensors are accurately capturing the information about the world, various mechanical components (such the steering wheel) are functional, the weather condition is good enough to yield accurate camera images, and the driver is attentive and ready to take over in case of a faulty sensor. If any of these assumptions fails to hold, a collision might occur; it would not matter how well tested or verified the control software is against its specification. 

Missing or incorrect domain assumptions are a major factor contributing to software failures. A National Academies study on software dependability, for example, documented that only a small percentage of software accidents (3\%) is due to bugs in software; the rest are caused by poor understanding of requirements and usability issues~\cite{NASDependability}. For another example, in a well-studied accident, Lufthansa Flight 2904 in 1993 overran a runaway during landing, resulting in 2 fatalities and 56 injuries. The cause was determined in part to be an incorrect assumption made by the software about the condition under which the thrust-reverse system was to be activated: the assumption was that the plane would be on the ground when its wheels are turning above a certain speed, which turned out to be violated when a wet runway caused the wheels to aquaplane~\cite{WarsawReport}. 

Domain knowledge will continue to play an important role in AI-driven systems. In fact, given its reliance on data, AI introduces additional types of assumptions on the world (such as data drifts) that must be carefully considered beyond those in traditional software systems~\cite{KastnerBlog}.

\subsection{Software engineering relies on tacit knowledge for problem understanding and solution design}

Tacit knowledge—knowledge that resides in an engineer or a stakeholder’s head and hence is undocumented—plays a crucial role in software development, especially in early phases such as problem understanding and exploration of design solutions. As we argue in Section 4.3, these tasks are unlikely to ever be amenable to complete automation, due in part to their dependence on such tacit knowledge. 

Understanding the domain—identifying relevant stakeholders in the world, eliciting requirements and domain assumptions, and validating these assumptions—is a challenging task in part because much of the knowledge resides in a domain expert or stakeholder’s head and often remains undocumented. Within the requirements engineering community, it is well-known that stakeholders often have a difficult time articulating what they actually want from a software product before they use it~\cite{requirements_elicitation}. Even in well-established organizations like NASA, important domain knowledge is informally “passed down” from one engineer to another over time~\cite{NASA-tacit-knowledge}. Failure to identify or communicate tacit knowledge has contributed to a number of well-known incidents in engineering systems, such as the infamous Chernobyl disaster, where important details about the nuclear reactor design were not properly communicated to the plant operators~\cite{chernobyl-accident}.

Beyond understanding the problem domain, the tasks of designing, implementing, and validating the software also involve a great deal of tacit knowledge. A software product is an embodiment of a set of design decisions that are made throughout a development cycle. But the final product does not capture a plethora of knowledge that is produced for decision-making—design alternatives considered, evaluation of trade-offs among them, and the rationale for the decisions made. This knowledge is often undocumented, difficult to recover from code, and ultimately lost when people with critical tacit knowledge leave an organization. It is, for example, widely believed that maintaining and evolving a software system becomes significantly harder when an engineer with an intimate knowledge of the system design leaves a project (a phenomenon also known as truck factor~\cite{AgilePatternsBook}). The role of tacit knowledge in software testing and validation has also been well studied, including the importance of personal experience and knowledge in tasks such as test development, fault localization, and bug fixing~\cite{ItkonenTSE13}. 

\subsection{Software engineering must also engage with systems engineering and the social context of the delivered software system}

In discussing design problems that arise in societal settings, Rittel and Webber identified the mismatch between the nature of these cybersocial problems and typical analysis strategies such as those common in software design~\cite{Wicked}.  They observed that each such problem is unique, and they described the difficulty of even getting a problem definition that satisfies numerous stakeholders with competing interests, and how this means that there’s not a common process for solving these ``wicked'' problems, and that there’s no criterion for success.

In a recent essay in NAE Perspectives, Madhavan argues (of engineering in general) that engineers need to develop a full systems engineering sensibility. He invokes wicked problems, then says “Engaging with wicked systems requires more than good intentions, creativity, and expertise. We need a communal code of conduct—or in an engineering sense, a concept of operations to train and treat our approaches (including, especially, education) to gain greater improvements. To do so, we need an engineering vision for civics and a civic vision for engineering. … Engineering is a carrier of history, simultaneously an instrument and the infrastructure of politics. It’s among the oldest cultural processes of know-how, far more ancient than the sciences of know-what. And through engineering, civics can gain a more structured, systemic, and survivable sense of purpose.” ~\cite{NatAcadEngr}

Software engineering has an obligation to foster design and development of systems that address the needs of the organizations and people they are intended to serve. Doing so requires an expansive vision of the criteria for success.

\section{SE has a long history of embracing radical new ideas from AI}

We are now said to be in the throes of an AI revolution, with a flood of technologies based on sophisticated statistical inference from large datasets that will challenge the old technologies. We are indeed seeing powerful new technology, but this is not the first time that AI has brought radical ideas to software.

In fact, AI has long been a source of new programming and software development techniques that initially seem radical or impractical but acquire respectability and are eventually adopted —with their origins in AI forgotten.  Many features of modern software development originated in AI and were assimilated into programming languages and techniques. This has been a process of evolution, not revolution.

\subsection{AI is an incubator for programming ideas}

Let’s look at just a few of the ideas that have made their way from the AI community to mainstream  programming languages and software development~\cite{Myths}. They range from low-level data structures to design approaches to broadening the conceptual basis of software.

List processing languages originated in the AI community well over half a century ago, introducing linked list structures~\cite{NewellShaw57} and functional programming~\cite{McCarthy60}. To the programming community, giving away half of memory—in a time when 32K of 36-bit words was a big machine—was at first shocking, but the utility of the representation established itself within a few years. Functional programming’s strong commitment to composition of pure functions and immutability was likewise a new mindset for mainstream programming, which was accustomed to direct manipulation of state.

AI also brought new design approaches to software. Programming methods were originally rooted in writing complete solutions to well-specified problems.  AI, on the other hand, recognized both algorithmic approaches (e.g., optimization) for well-specified problems and heuristic approaches (e.g., generate-and-test, rule-based systems) for  ill-structured problems~\cite{Simon96}. These ill-structured problems often involved aspirations that were not well defined in advance, and writing software was a way to understand what was actually achievable or desirable. For example, production systems emerged as a technique for implementing expert systems, which were typically developed incrementally by trial and error. The acceptability of partial solutions as a means of understanding shaping the objectives for a software system set the stage for exploratory programming, which is now common in SE~\cite{Myths}. 

A third way that AI contributions expanded the scope of software was the conceptual shift from computation over numbers and strings to a broader view of computing that included manipulating symbols as well. Early programmed systems computed with numeric values and strings.  AI introduced high-level symbolic representations of problems and logic for tasks such as symbolic mathematics, knowledge-based systems, theorem proving to support reasoning tasks, and decision making.  Early examples of symbolic processing include the Logic Theorist, which performed automated reasoning~\cite{LogTh56}, and Macsyma, a general purpose computer algebra system~\cite{Moses2012}. This richer semantic basis facilitated later developments in programming such as reflection, self-modification, and self-adaptation.
In its quest for solutions to problems that exemplify human intelligence, AI had to attack complex ill-structured problems at a time when the prevailing sensibility of the programming language research community was still focused on specification, sound languages, and proving correctness. The associated challenges led to innovations that became integrated in software engineering.  In addition to the examples described here, software techniques originating in AI include garbage collection, logic programming, backtracking, reflection , and some types of search.

\subsection{Generative AI is another disruptive new idea from AI}

Generative AI is the new kid on this block. Its claim to disruptive power is rooted in its fundamentally statistical nature, and the challenges it presents to SE arise from this.  The principal challenge, of course, is the large amount of data used to train the models, with the associated challenges of curating the data.  In addition, there are concerns about opacity, nondeterminism, lack of specifications, correctness, and reliability.  Let’s consider each of these in turn.

The most obvious novel aspect of generative AI is its collection, curation, and processing of very large datasets. The training sets rely on the automated collection, classification, cleansing, versioning, and analysis of data from often-uncontrolled sources.  This data contains errors of various kinds: it may be incorrect, ill-formed, inconsistent, malicious, ambiguous, or biased.  However, ordinary database systems also have large, rich, highly structured data managed separately from the code, and they deal with errors and inconsistencies in the data. SE and database systems have historically been largely independent, but database expertise can complement SE expertise to manage vast amounts of noisy data.

The other properties of AI’s data that are called out for concern also have analogs in established SE practice~\cite{Myths}:
\begin{itemize}
\item \emph{Opacity: The models underlying generative AI are opaque, and they can’t explain how their outputs are related to their inputs.}  However, most conventional software components are also opaque in practice. They are used without extensive documentation, and although we believe in principle that we could understand them by reading the code, in practice the code is often not available and, if it were, the task of understanding it would be daunting.
\item \emph{Nondeterminism and dynamically changing behavior: AI components are regularly retrained, so their behavior changes without notice.} However, nondeterminism is familiar to SE; for example, any system with physical components is nondeterministic, and any system involving human interaction receives unpredictable inputs. Further, web applications and ordinary components provided by third parties in the cloud, for example ThingOfTheDay-as-a-Service, also change behavior without notice. The sources of nondeterminsim may vary, but the techniques of dealing with it are well-established.   
\item \emph{Lack of specifications and correctness: Generative AI systems are intended to model whatever was in the input data.} If we could specify that precisely, we wouldn’t need the AI models; the objective is for the output to be sufficiently similar to the training data, but the criteria for sufficiency are imprecise. Without precise specifications, strict correctness would not be possible even if the software were algorithmic and deterministic. AI results  must be interpreted in the context of imprecise specifications, so strict correctness is not a reasonable goal.  However, many systems in the real world, especially socio-technical systems, also lack precise specifications, and we judge them on their fitness for purpose.  We return to this in Section 5. 
\item \emph{Reliability: Machine learning systems appear to be unreliable, partly because of the absence of specifications and because of hidden dependencies and model interactions~\cite{kastner2021}.} Incidental data in the training set (e.g., image backgrounds) may obscure essential data (e.g., foreground images). However, SE has long-established principles for developing reliable systems from unreliable parts with techniques such as modularization, redundancy, runtime checks, firewalls, fault tolerance, and so on that compensate for the vagaries of unreliable components~\cite{lyu1996}.  Feature interaction, in particular, has been studied for decades ~\cite{kastner2021}. Again, the sources of unreliability may be different, but SE has a rich portfolio of responses that can be adapted.
\end{itemize}

Our reasoning resembles that of Shapiro and Varian, who addressed claims that the new network and information network economy requires a completely new economics~\cite{InfoRules}. No, they said, “Technology changes, economic laws do not”.  That is, the durable principles of economics work just fine in the network economy, it’s just that the parameters are different. Applying that reasoning to SE, we should recognize the new technology and ask how the old principles, perhaps with modifications, still apply.  The unruliness of machine learning components may be greater than that of more conventional components, but we should expect to address them by evolving the techniques we’ve developed for more familiar software.

\section{SE will evolve to handle generative AI}

AI has the potential to contribute to SE in several ways, provided it can become trustworthy. What can happen if we don’t panic?
\begin{itemize}
    \item The level of abstraction in programming may rise again; this will not eliminate programming, but, as with previous programming language innovations, it will increase the leverage of each line of code.
\item Generative AI may lead to new sorts of software system architectures. This might take the form of new types of components and connectors, and it might take the form of variants on established components that respect the stochastic nature of generative AI.
\item High-level design, which is context-specific and relies on large amounts of tacit knowledge, is unlikely to be threatened. Similarly, aspects of development that rely on engineering judgment don’t lend themselves to stochastic generation from written texts. AI-based tools for some kinds of routine engineering may be able to provide recommendations but will require human supervision.
\item Development tools for vernacular programmers may improve, though they will likely require supervision.
\end{itemize}

In addition, formal methods are predicated on correctness as the gold standard, but much of the work in this area is intended for software artifacts with traditional logical semantics. It is not compatible with the probabilistic and stochastic nature of generative AI, which relies on syntax and surface analysis of data rather than underlying meaning. Indeed, that view of correctness falls short for many types of traditional SE, especially cyber-social systems and exploratory programming. We address this below in Section 5~\cite{PointCounter, Myths}.

\subsection{Generative AI may raise the level of abstraction of programming, but it won’t eliminate the jobs}

Generative AI already contributes to programmer productivity~\cite{mckinsey-study}. Much of the code written in day-to-day production is repetitive and stereotypical---the stuff of routine engineering, as discussed in Section 2. It is likely that the success of generative AI in routine coding will arise from the large body of similar code to use as training data.

As AI’s code suggestions become more reliable, we foresee these programming cliches becoming accepted programming abstractions (following the pattern of Section 2.1). After all, the best candidates for automation are things that are well-understood, tedious, and error-prone when left to humans.  This is not to say that generative AI will create novel abstractions, but rather that it may identify repetitive phrases that can be encapsulated and named.

Current concerns about AI eliminating all the programming jobs can be reframed as raising the level of abstraction at which we program, just as compilers and frameworks raised the level of abstraction. Those technologies eliminated some jobs for programming at the machine level  while adding jobs for programming with higher-level languages.  However, the real engineering of software takes place far above the code, where judgment and tacit knowledge rule (and where there aren't zillions of pages of similar stuff written down, so statistical similarity doesn't have much leverage). 

We began by observing that the prospect of generative AI taking over the programming task generates cries of alarm. However, current industry estimates still show a severe shortage of programmers---with a shortage of qualified programmers rising to 4 million by 2025~\cite{devnet24}. In response, the industry has turned to “low-code” and “no-code” development environments in which users with little formal coding experience use visual interfaces to create apps~\cite{IBMlocode}. Gartner predicts that the low-code market, already strong, will continue to grow~\cite{gartner22}.  It’s reasonable to assume that generative AI will improve the current tools for these routine tasks. This seems more likely to alleviate the shortage of programmers than to create massive unemployment.

More broadly, Autor studied the history of workplace automation in the 20th and 21st centuries~\cite{Autor15} and reported, “Automation does indeed substitute for labor—as it is typically intended to do. However, automation also complements labor, raises output in ways that lead to higher demand for labor, and interacts with adjustments in labor supply. …  journalists and even expert commentators tend to overstate the extent of machine substitution for human labor and ignore the strong complementarities between automation and labor that increase productivity, raise earnings, and augment demand for labor”. With respect to AI he argues that “the interplay between machine and human comparative advantage allows computers to substitute for workers in performing routine, codifiable tasks while amplifying the comparative advantage of workers in supplying problem-solving skills, adaptability, and creativity.“

\subsection{Generative AI will expand the vocabulary of software architectures}

SE already deals with different kinds of components (such as objects, procedures, filters, and processes, and microservices) and with connectors (such as method invocation, procedure call, data flow, and implicit invocation); this is the vocabulary of software architecture. AI components will surely have different interfaces and different protocols for interacting with other components. This will require evolution of current techniques for software architecture, but there’s no indication that this will be an insurmountable challenge.

On the surface, LLMs may be perceived as a type of software-as-a-service (SaaS) that receives an input (a natural language prompt) and generates an output (a response), to be invoked from and integrated into part of a software system. Successful, long-lasting services share some key properties: They are predictable (in that, if they are invoked by the client repeatedly with the same input, they show the consistent behavior), reusable (they can be used in different contexts and produce an expected behavior), and reliable (they produce the behavior that is promised to the client through their interface). Due to their inherent statistical nature, however, the current state-of-the-art LLMs do not provide these properties out-of-the-box: The same input might produce vastly different responses. For a given task (e.g. program generation), while producing reliable responses for one type of input (e.g. a popular language such as Python), it may fail to do so for others (less well-known languages, such as Lua). LLMs will likely continue to suffer from the problem of so-called hallucinations\footnote{The current fashion for calling generative AI's errors “hallucinations” is appalling. It anthropomorphizes the software, and it spins actual errors as somehow being idiosyncratic quirks of the system even when they're objectively incorrect. Moreover, WebMD’s advice about hallucinations is “If you or a loved one has hallucinations, go see a doctor.”  \url{https://www.webmd.com/schizophrenia/what-are-hallucinations}
}, where they generate plausible-sounding but incorrect responses. Finally, as these models are re-trained and updated from time to time, there is no guarantee that their original behaviors will be preserved or that their performance will strictly improve.   

To manage the inherent uncertainty in LLMs and build reliable systems around them, additional engineering effort is likely to be necessary. This may involve, for example, (1) devising an interface that clearly states assumptions on the input and expectations on the resulting behavior of an LLM, (2) deploying  guardrails to prevent LLMs from producing risky, unexpected behaviors, and (3) developing an approach to check the quality of an output generated by the LLM (using retrieval augmented generation (RAG)~\cite{rag}, for example). Methods for identifying hazards and dealing with failures, based on well-established methods from reliability and safety engineering~\cite{Leveson,lyu1996}, will be a critical part of this process. Existing architectural styles, connectors and patterns will need to be adapted or new ones developed, for structuring a system as a composition of both traditional software and LLM-based components. Techniques for explicitly representing and reasoning about uncertainty in programs (such as probabilistic programming~\cite{probabilistic-programming}) may also play a role.

In summary, a significant amount of effort will be needed to turn LLM-based components into ``plain old software"---those that we can rely on as being predictable, reusable, and reliable~\cite{POSW}.

\subsection{AI does not threaten higher level concerns like requirements engineering, design, and reliability}

Generative AI has potential to be an effective aid for tasks such as requirement elicitation, design, and reliability engineering, but it is unlikely to completely automate or replace them. These tasks are highly-context dependent, and they demand flexibility, judgment, common sense, and a great deal of tacit knowledge, as discussed  in Section 2.4.

Researchers are exploring ways to use LLMs for these tasks—for example, to synthesize formal software specifications from natural language prose~\cite{nl2spec}, validate the consistency of requirements~\cite{Breaux24RE}, and identify missing requirements~\cite{LuitelHS23}. Ultimately, however, requirements must be validated for alignment with the business constraints and stakeholders’ needs from a specific problem context—a process that involves human interactions and is not currently automatable by LLMs. 

Design space exploration is another type of activity that may be supported, but is unlikely to be replaced, by generative AI. Software engineers are faced with a wide range of design decisions, where no single solution is superior to the others and where they present trade-offs among different quality attributes (e.g., performance, usability, security, reliability, etc.) ~\cite{DesignSp}. LLMs may one day be useful for generation of well-known design options, or even evaluation of an individual option with respect to a quality attribute. But ultimately it will be up to the engineer to navigate a space of possible solutions, apply sound engineering judgment, and arrive at a decision about which solution should be selected for each particular project. 

\subsection{AI may improve support for vernacular programmers}

Vernacular developers---people who are not highly trained programmers or software engineers  but who create and adapt software for their own goals---vastly outnumber trained programmers.  They are often professionals in their own fields, with principal responsibilities and training in those fields, and they develop software to solve problems within their particular professional contexts.  They are also often hobbyists who write software in pursuit of their personal interests~\cite{Myths}.

Vernacular developers have been poorly served by the computing community for many years. Much of their computing is stereotypical or exploratory, and they tend to judge the adequacy of software by examining its behavior, not by the more stringent validation standards of professional software engineers. To the extent that AI raises the level of abstraction, they may be able to describe their programs in terms closer to their own problem domain and benefit from AI. In addition, the potential of generative AI to automate other development tasks such as version control (e.g., Git), system build, testing, and deployment~\cite{autodev} may further improve the support for vernacular developers.  

The low-code movement improves support for those whose needs align with the low-code tools by allowing them to describe their intentions at a level above code. They can look forward to support from generative AI, at least for the tasks that are common enough for the code patterns to be well-established.

\section{SE needs to re-think its concept of correctness; generative AI may force SE to do this}

Most significantly, AI may finally force SE to reconsider what “correctness” means. As we observed in Section 2, SE’s origin story was about specifications and correctness. Although SE’s fixation on formal correctness has softened somewhat over the years, there is still a widespread cultural mindset that specification and correctness are major objectives, even if full verification is unachievable. Recall, though, that program verification can only show the consistency between the code and some formal specification. It cannot show that the formal specification actually captures the client’s intentions, and any properties that can’t be expressed in the specification language are unverifiable, at least in the formal system.

This traditional view of correctness is inadequate to serve many conventional software systems. Cyber-physical systems have intrinsic uncertainty arising from mechanical linkages. Software “in the cloud” may update without notice, thanks to modern continuous integration and deployment strategies. Exploratory programming co-evolves understanding the requirement with developing the software, so there may be no formal specification, especially at the outset. Societal problems don’t even have consensus definitions, let alone specifications~\cite{Wicked}. Vernacular developers often do not include mathematical models in their understanding of their problems. Even if a formal specification is available (and it’s costly to produce one), proving correctness of the specified properties can be very expensive.

As to validation of generative AI systems, the complex models that power these systems were trained using large bodies of examples drawn from real-world sources with the objective of being able to produce output that is consistent with the sources.  This essentially probabilistic reasoning isn’t the same thing as rigorous symbolic reasoning, whether it be formal verification or code analysis. It simply isn’t. We should recognize the difference and address it head-on.

Let’s consider other, more nuanced, ways to think about whether software is acceptable. What might be a more appropriate view of software quality than strict correctness?  We can draw on some examples:
\begin{itemize}
\item Assurance cases for compliance with safety standards rely on combinations of test cases, proofs, and human judgment.
\item The current view of specifications as formal, static, and complete could be replaced with a model that they are heterogeneous, evolving, and incomplete, and they capture the  most appropriate information available at a given time.  We call these credentials~\cite{Credentials1,Credentials2}.
\item In civil and mechanical engineering, designs incorporate generous safety factors to guard against uncertainties such as variability of materials and uncertainties in the design.
\item A bit farther afield, city building codes provide safety-motivated constraints on construction that can be followed by ordinary handyfolks as well as professional architects and contractors. These constraints identify certain tasks that must be done by licensed professionals (electrical and gas work, for example).
\end{itemize}

To do this, let’s consider some alternative objectives to this notion of correctness for the acceptability of software:  fitness for task, trustworthiness, and statistically likely outcomes.

\begin{figure*}[!t]
\includegraphics[width=0.7\textwidth]{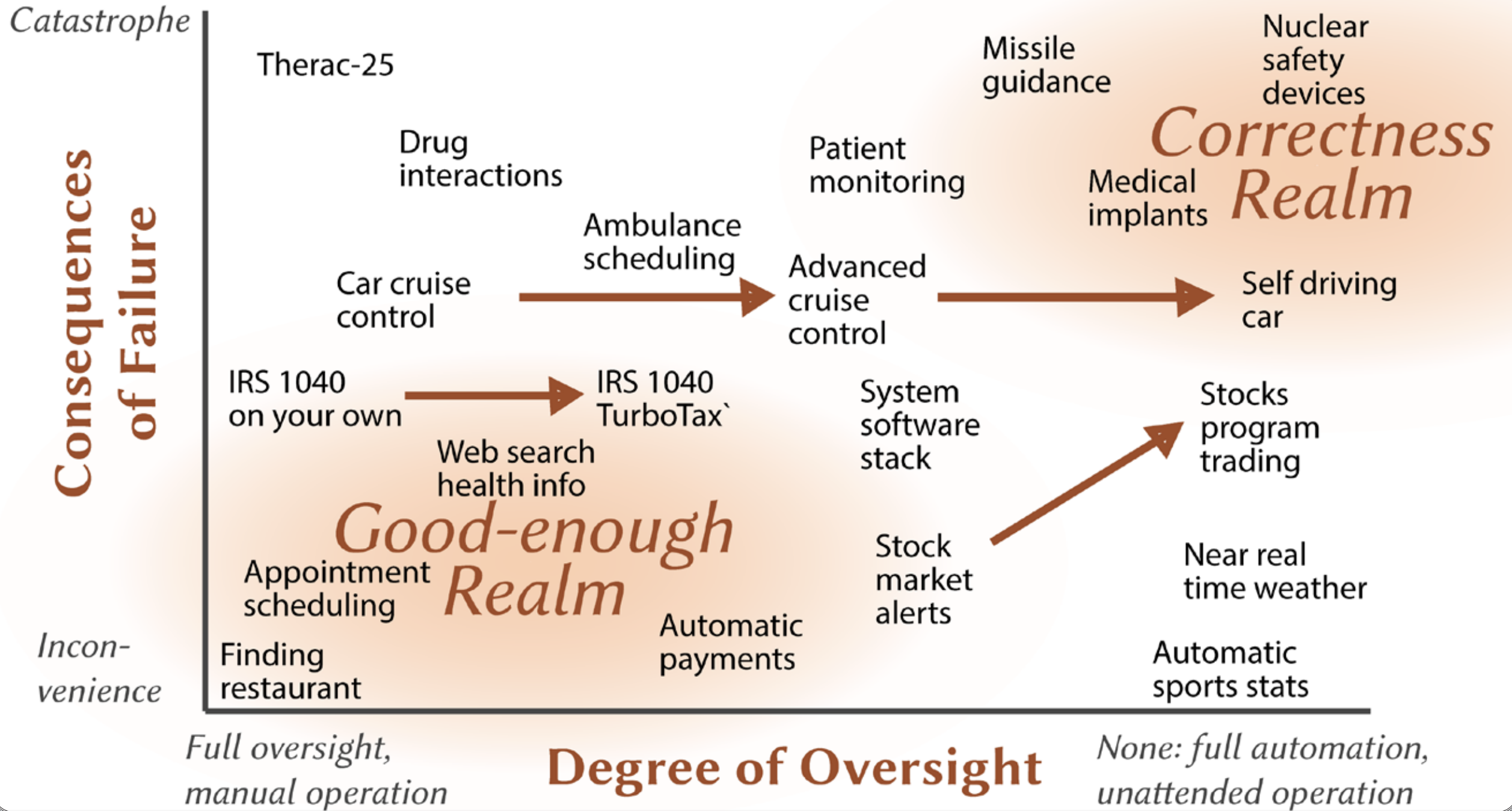}
\centering
\caption{How consequences and human oversight affect the level of validation needed~\cite{Myths}}
\label{fig:correctness-realm}
\end{figure*}

\subsection{Correctness vs fitness for task}

We suggest that “fitness for intended purpose” is a better goal than formal “correctness”. Traditionally, establishing correctness relies on the existence of a formal specification that unambiguously captures the intended behavior of a piece of software, or at least a subset of that. Given that specification, verification shows consistency of code and specification.

“Fitness for purpose” retains the ability to mean traditional correctness for critical systems where formal proof is the best way to establish fitness. However, it also recognizes the legitimacy of more informal ways to show that software is “good enough” for its purpose. This echoes Alexander’s emphasis on the need for design to establish fitness between the form and its context~\cite{alexander-synthesis}. Figure~\ref{fig:correctness-realm} illustrates this by locating applications in a space whose axes are the consequences of failure and the degree of human oversight. This reframing of “correctness” requires thoughtful consideration of how good is “good enough” for each application, in its operational context. 

Thinking about what’s “good enough” also shows where it’s worth paying the premium for formal specifications and correctness (where human oversight is low and consequence of failure is high) and where the error rates of generative AI can be tolerated (where human oversight is high and consequences of failure are low).  This is consistent with the robustness rule of thumb: If consequences of failure are high, prevent the problem and validate the code; if consequences of failure are low, it’s OK to remediate the problem after detecting it~\cite{Robust}.

Expressing the developer’s intent through natural language prompts will likely emerge as the dominant paradigm for writing software in the future. The code generated by AI tools must eventually be validated against the original intent, which, unfortunately, will rarely be available as a formal specification and will likely be imprecise. For certain software-driven tasks (e.g., exploratory programming), it may be unclear whether a specification exists at all. Instead, programs constructed in this manner will be associated with an abundance of informal, semi-structured natural language prose. In place of the traditional formal notion of correctness, a notion of fit or consistency between the generated artifact and the developer intent (reflected through the prose) will be needed, along with methods for checking these properties.

\subsection{Fitness for explicit task vs trustworthiness}

Even the reframing of “good enough” is not good enough.  For software embedded in a social context, we need to pay more attention to whether the intent itself is appropriate. For example, an LLM that delivers results consistent with its training data might arguably be correct (i.e., it replicates human decisions), but if the training data is biased relative to societal norms, the results would be unacceptable in the context of a fair decision system (i.e., the human decisions it’s replicating are biased or discriminatory).   Sociotechnical systems that attempt to solve wicked problems are especially sensitive to the difficult-to-define criteria for success, which may have complex ethical, cultural, and political components ~\cite{Wicked}.

The National Institute of Standards and Technology (NIST) identifies the essential building blocks of AI trustworthiness as~\cite{NIST}:  
\begin{itemize}
    \item Validity and Reliability
\item Safety
\item Security and Resiliency
\item Accountability and Transparency
\item Explainability and Interpretability 
\item Privacy
\item Fairness with Mitigation of Harmful Bias
\end{itemize}
Some of these elements can be assessed quantitatively, either directly or by proxies. Others may be measured qualitatively, or they may reflect perceptions of the user.

Trustworthiness goes beyond adherence to a specification. It depends on context of use and individual expectations; that is, it depends on the perception of the user as well as the intent of the developer, so a given system might be trustworthy in some contexts but not others~\cite{SEI2023}.

\subsection{Trustworthiness vs statistically likely outcomes}

Generative AI is fundamentally based on sophisticated statistical analysis of existing artifacts.  Consequently, its results are reflections of statistical likelihood, not of validated semantic models---they are not sound in the usual formal sense. It follows that as the underlying statistical model is updated, the results may also change.  Achieving trustworthiness requires some kind of supervisory process. Although recent innovations such as RAG may provide some level of  confidence about model results in certain scenarios by referencing an external knowledge base, human oversight will likely remain indispensable for the foreseeable future. 

There is a great deal of regularity and stereotypical phraseology in what people write, both in natural language and in programming languages. Indeed, automated program repair exploits this regularity to find fixes~\cite{HindleBSGD12}. Reifying this regularity contributes both to natural language cliches and to software abstractions. 

But similarity in phrasing is not a substitute for accuracy, let alone for semantics. A sentence whose structure matches human writing or speech patterns may be syntactically plausible, but there’s not much assurance that the word choices in any particular instance match underlying reality.  We see this in full flower in the numerous examples of papers and legal briefs in which genAI cites nonexistent material: The sentence pattern of the citation is conventional, but the precise thing to cite is elusive, so the citation may simply be invented. This is consistent with experiments that compared genAI performance on familiar tasks with performance on unfamiliar formulations of those tasks. Performance is poorer on unfamiliar tasks, suggesting that current LLMs rely more on narrow non-transferable procedures than on abstract task-solving skills~\cite{wu24}.

Generative AI is most likely to be useful in applications where its benefits outweigh the risks stemming from the uncertainty of its behavior (the “good-enough” realm in Figure~\ref{fig:correctness-realm}), or where an automated or human-guided mechanism validates its output. Examples of the former can be found in creative endeavors that involve production of images, text, or videos; they may also arise in business applications where responses to frequent custom queries may be generated by AI, with a human operator on standby to intervene when the AI agent’s result is unsatisfactory. Code generation is an example of the latter; it has been observed that programmers who use an LLM  for coding tasks often end up manually inspecting the generated code to ensure that it conforms to their intent~\cite{copilot-study}. 

One vernacular developer with decades of experience reports using Copilot~\cite{GithubCopilot} in PHP programming; he has done so for long enough that Copilot has seen a lot of his code~\cite{Weil24}. When he types a line of code, Copilot will suggest one or two lines that might follow, which he can accept by hitting tab or ignore. He accepts the suggestion intact 50\% of the time, and 20\% of the time he accepts it and does additional tweaking. The suggestion is obviously not useful (wrong or irrelevant) 30\% of the time.  As one outstanding example, he defined a SQL query, and Copilot suggested a dozen next lines that invoked the query, iterated over the records, and did for each record an operation similar to what he had done before in a similar situation. He was amazed.

A recent survey of software developers~\cite{Stackoverflow24} found that over 60\% of developers are using AI tools, mostly for writing code and searching for answers. However, 66\% of developers don’t trust the output or answers produced by AI tools; 63\% were concerned that the AI tools lack the context of the codebase, internal architecture, and/or company knowledge; and 45\% believe AI tools are bad or very bad at handling complex tasks.

While generative AI may someday have a role to play in systems with a higher level of criticality, its use for safety-critical functions is likely to be limited unless proper safeguards can be put in place. In addition to “hallucinations,” LLMs are also susceptible to misuse, bias, and privacy issues. Just as  the safe deployment of self-driving vehicles is taking longer than predicted by many, deploying generative AI in a safe, trustworthy manner will become viable only after breakthroughs in more reliable reasoning about AI technologies and/or methods for developing guardrails around these models emerge.

In fact, generative AI has a long way to go.  Recently a hallucinated package name was instantiated by a real person and downloaded many times. If the real person had instantiated it maliciously, important software (including AliBaba) could have been compromised.  This speaks not only to the weakness of generative AI but also the need for users to exercise great care ~\cite{badpackage}.

\subsection{Wider, societal risks of generative AI }

Software engineers must also carefully consider the long-term impact of deploying an AI-driven system on the real world. LLMs are developed and trained based on data collected and curated from the world, which, in turn, may be influenced and shaped by these models. This type of feedback loop can cause harmful effects in some of the most important aspects of our society, such as education, healthcare, employment, justice, and social media~\cite{oneil2016}. Numerous examples of societal bias that are reflected and reinforced by LLMs are well documented~\cite{llm-risks}. Although these are not new problems, they are likely to be exacerbated by the rapidly increasing use of generative AI technologies in socio-technical systems. 

The long-term effect of LLMs on creativity of human developers should also be considered. LLMs provide easy access to sample design and coding solutions, and they facilitate rapid prototyping of an application. However, when developers become over-reliant on these tools, they may become less inclined to explore alternatives beyond those that are generated for them by AI. These “standard” solutions would be collected for retraining the LLMs, which, in turn, would be again used to generate and introduce similar code into the world. When this reinforcing feedback loop remains unattended, it might gradually and adversely impact the diversity of software solutions that are used and deployed in the world. This type of problem, also called model collapse, has already been shown to be a rising problem in modern LLMs~\cite{shumailov2024,guo2024}. 

Similarly, although generative AI has the potential to make programming accessible to a wider group of the population and support vernacular developers in their daily tasks, we must also prepare for ways to mitigate possible risks. These risks include at least proliferation of low-quality, poorly tested code on app stores and open-source ecosystems, intellectual property and privacy violations, and exploitation of the generative technologies for malicious purposes (e.g. generation of deep fakes and misinformation).

On a more positive note, the automation of repetitive development tasks may free up time for more creative aspects of software development and ultimately help enable a more diverse range of novel software products than are possible today. To achieve this outcome, however, we believe that developers should make a conscious effort (and be encouraged by their organizations) to balance their use of generated artifacts with a deliberate consideration of the problem context, stakeholders’ needs, and possible design options.

\section{How, then, should SE engage with generative AI?}

Generative AI is now eagerly inflating our aspirations, but its capability is not yet trustworthy and robust enough to be part of the stable core of SE methods. AI is already demonstrably useful under careful supervision, and we can expect its utility for routine programming tasks to improve quickly. It may serve as an assistant for higher levels of design, but the tacit knowledge that drives those activities is largely inaccessible to AI training sets, so experienced software engineers will retain the initiative.  For the foreseeable future, though, we expect AI outputs to best be treated as suggestions for review.

In the 1970s there was considerable attention to “the software crisis” and to “solving the software problem,” as if there could be a clear solution~\cite{NATO68}. However, our aspirations for software will always exceed our current capability to produce the software; that is, we are always reaching for software systems in an aspirational cloud outside our stable, reliable core capabilities. As our capabilities increase, our imagination expands our aspirations at least as fast.  So we’ll never fully “solve” software; the best we can hope for is to steadily grow the stable core capabilities in order to expand the envelope of things we still find challenging.

Viewed in this light, software creation with generative AI is still in the aspirational cloud outside the stable core.  That means we should use it carefully, with strong human oversight, automated checking and strong guardrails. As we gain more experience, we can hope that it earns trust worthy of core capabilities.

Finally, the essentially probabilistic and stochastic nature of generative AI does not lend itself to formal specification and proof, so validation of software involving generative AI requires reasoning and analysis appropriate to those underlying mathematical models.  But a great deal of conventional software development does not lend itself to formal specification and proof, despite being rooted in symbolic and deterministic models. Perhaps the challenge posed by AI will finally bring SE to confront the need to dramatically rethink the idea of “correctness.”

\section*{Acknowledgements}
We appreciate suggestions, discussion, and feedback from colleagues, including the CCRC/IP-ASIA discussion group of Keio University, Irving Wladawsky-Berger, Rodney van Meter, Wendy Grossman, Richard P. Gabriel, Bill Scherlis, Nancy Mead, Carol J Smith, Marian Petre, Ipek Ozkaya, and three anonymous Onward! Essays reviewers.

This essay sets the software engineering context by synthesizing and refining several long-standing positions. These include the view of software engineering as an engineering discipline \cite{Prospects} \cite{Manifesto}, the use of abstraction scale as a proxy for progress \cite{Abstraction}, and the case for software engineering adapting to embrace AI as laid out in \cite{Myths} and elaborated in \cite{PointCounter}.  

The discussion of modern AI draws  on experience in developing and teaching a course on software engineering for AI-based systems (https://mlip-cmu.github.io).

This work was supported by the National Science Foundation Award 
No. 2144860 and the A. J. Perlis Chair of Computer Science at Carnegie Mellon.

\bibliographystyle{ACM-Reference-Format}
\bibliography{refs}

\end{document}